\documentclass{elsart}
\usepackage{times}
\usepackage{courier}
\usepackage{algorithm}
\usepackage{algorithmic}
\usepackage{natbib}
\usepackage{epsfig}
\usepackage{amssymb}

\def\dj{\hbox{d\kern-0,347em \vrule width0,3em height1,252ex
depth-1,21ex \kern0,051em}}
\begin{document}

\begin{frontmatter}

\title{An Efficient Algorithm For Simulating Fracture Using Large Fuse Networks}

\author[1]{Phani Kumar V.V. Nukala}
\author[1]{Sr{\dj}an \v{S}imunovi\'{c}}
\address[1]{Computer Science and Mathematics Division, 
Oak Ridge National Laboratory, Oak Ridge, TN 37831-6164, USA}

\begin{abstract}
The high computational cost 
involved in modeling of the progressive fracture simulations using large discrete lattice 
networks stems from the requirement to solve {\it a new large set
of linear equations} every time a new lattice bond is
broken. To address this problem, 
we propose an algorithm that combines the multiple-rank sparse Cholesky downdating 
algorithm with the rank-p inverse updating algorithm based on the 
Sherman-Morrison-Woodbury formula for the simulation of progressive
fracture in disordered quasi-brittle materials using discrete lattice
networks. Using the present algorithm, the computational complexity of solving the new set of linear
equations after breaking a bond reduces to the same order as that of a simple {\it
backsolve} (forward elimination and backward substitution) {\it using the
already LU factored matrix}. That is, the computational cost is $O(nnz({\bf L}))$, 
where $nnz({\bf L})$ denotes the number of non-zeros of the 
Cholesky factorization ${\bf L}$ of the stiffness matrix ${\bf A}$. This algorithm using the direct sparse
solver is faster than the Fourier accelerated preconditioned conjugate gradient (PCG) 
iterative solvers, and eliminates
the {\it critical slowing down} associated with the iterative solvers
that is especially severe close to the critical points.
Numerical results using random resistor networks substantiate
the efficiency of the present algorithm.
\end{abstract}

\begin{keyword}
\PACS 62.20.Mk \sep 46.50.+a
\end{keyword}

\end{frontmatter}

\section{Introduction}
Progressive damage evolution leading to failure of disordered
quasi-brittle materials has been
studied extensively using various types of discrete lattice models
\cite{deArcangelis85,sahimi86,duxbury86,duxbury87,hansen001,herrmann90,sahimi98,chakrabarti}.
Large-scale numerical simulation of these lattice networks in which the damage is 
accumulated progressively by 
breaking one bond at a time until the lattice system falls apart has often been hampered due to 
the fact that a new large set of linear equations has to be solved everytime a lattice bond 
is broken. Since the number of broken bonds at failure, $n_f$, increases with increasing 
lattice system sizes, $L$, i.e., $n_f \sim O(L^{1.7})$, numerical simulation of large 
lattice systems becomes prohibitively expensive. Furthermore, in fracture
simulations using discrete lattice networks, ensemble averaging of
numerical results is necessary to obtain a realistic representation of
the lattice system response. This further increases the computational cost 
associated with modeling fracture simulations in disordered quasi-brittle materials 
using large discrete lattice networks.

\par
Fourier accelerated PCG iterative
solvers \cite{batrouni86,batrouni88,batrouni98} have been used in the
past for simulating the material breakdown using large lattices. However, these
methods do not completely eliminate the {\it critical slowing down}
associated with the iterative solvers close to the critical point. 
As the lattice system gets closer to macroscopic fracture, the condition
number of the system of linear equations increases, thereby increasing
the number of iterations required to attain a fixed accuracy.  This
becomes particularly significant for large lattices.
Furthermore, the Fourier acceleration technique is not effective when
fracture simulation is performed using central-force and bond-bending
lattice models \cite{batrouni88}.

\par
This study presents an algorithm  
that combines the multiple-rank sparse Cholesky downdating scheme with 
the rank-p inverse updating scheme of the stiffness matrix, which effectively reduces   
the computational bottleneck involved in 
re-solving the new set of equations after everytime a bond is broken.
In this paper, we consider a {\it random threshold} model problem, where a lattice consists of 
fuses having the same conductance, but the bond breaking thresholds, $i_c$, 
are based on a broad (uniform) probability distribution, which is constant between
0 and 1. This relatively simple model has been extensively used in the literature 
\cite{deArcangelis85,sahimi86,duxbury86,duxbury87,hansen001,herrmann90,sahimi98} for 
simulating the fracture and progressive damage evolution in brittle materials, and provides a 
meaningful benchmark for comparing different algorithms. 
A broad thresholds distribution represents large disorder and
exhibits diffusive damage leading to progressive localization, whereas a very narrow
thresholds distribution exhibits brittle failure in which a single crack propagation
causes material failure. Periodic boundary conditions are imposed in the horizontal direction to simulate
an infinite system and a constant voltage difference (displacement)
is applied between the top and
the bottom of lattice system. The simulation is initiated with a triangular lattice of intact
fuses of size $L \times L$, in which disorder is introduced through random breaking thresholds. The voltage
$V$ across the lattice system is increased until a fuse (bond breaking) burns out.
The burning of a fuse occurs whenever the electrical current (stress)
in the fuse (bond) exceeds the
breaking threshold current (stress) value of the fuse. The current is redistributed
instantaneously after a fuse is burnt. The voltage is then gradually increased until
a second fuse is burnt, and the process is repeated.
Each time a fuse is removed, the electrical current is redistributed and hence it is 
necessary to re-solve Kirchhoff equations to determine the current flowing in the
remaining bonds of the lattice. This step is essential for determining 
the fuse that is going to burn up under the redistributed currents. Therefore, 
numerical simulations leading to final breaking of lattice system network are very 
time consuming especially with increasing lattice system size. 

\subsection{Summary of the Proposed Algorithm}
The algorithm presented in this paper reduces the computational complexity of obtaining the
solution ${\bf x}_{n}$, after the $n^{th}$ bond is broken, to a backsolve using the
already existent factorization of the stiffness matrix ${\bf A}_{m}$, and
$p = (n-m)$ vector updates. The algorithm is
based on the well known Shermon-Morrison-Woodbury formula \cite{golub96}
for obtaining
the inverse of the new stiffness matrix ${\bf A}_{n+1}^{-1}$
(after the $(n+1)^{th}$ fuse is burnt)
from the old stiffness matrix inverse ${\bf A}_{n}^{-1}$ through a rank-one update.
Infact, the algorithm is such that if the inverse of the lattice stiffness
at any stage ($m = 0,1,2 \cdots$) of analysis ${\bf A}_{m}^{-1}$ is available, then
all subsequent analysis involving ($n = m+1,m+2, \cdots$) burnt fuses can be
carried out using $p = (n-m)$ vector updates. However,
since most often the inverse of the stiffness matrix is
rarely ever explicitly calculated, the algorithm additionally
requires a backsolve
using the already existent factored matrix ${\bf A}_{m}$.
The backsolve operation is further simplified by the fact that it is performed
on a trivial load vector and hence the solution can be obtained easily.

\par
Based on the above description of the algorithm presented in this paper, given
the factorization of the matrix ${\bf A}_{m}$, the computational
cost involved in all the subsequent steps ($n = m+1,m+2, \cdots$) is a backsolve using the
already factored matrix, and $p = (n-m)$ vector updates. The computational complexity of
the backsolve is $O(nnz({\bf L}_m))$, where $nnz({\bf L}_m)$ denotes the number of non-zeros of the
Cholesky factorization ${\bf L}_m$ of ${\bf A}_{m}$. The computational complexity of
$p$ vector updates is $O(p~n_{dof})$, where $n_{dof}$ denotes the
number of degrees of freedom in the system. As $p$ increases, it is possible that
the computational cost associated with the $p$ vector updates exceeds the cost involved in the
factorization of the matrix ${\bf A}_{n}$. Under these circumstances, it is advantageous to
obtain the factorization ${\bf L}_n$ of the new stiffness matrix ${\bf A}_{n}$, and use this
${\bf L}_n$ for all the subsequent backsolve analysis steps, until the computational cost
associated with the vector updates once again exceeds the stiffness factorization cost.
Using the algorithm presented in the paper, it is not necessary to
re-factorize the new stiffness matrix ${\bf A}_{n}$. Instead, we
adopt the multiple-rank update of the sparse Cholesky factorization algorithm
\cite{tdavis1,tdavis2} for
updating the ${\bf L}_m \rightarrow {\bf L}_n$. This multiple-rank update of ${\bf L}_m$ to
obtain the new factorization ${\bf L}_n$ is
computationally cheaper compared with the direct factorization ${\bf L}_n$
of the new stiffness matrix ${\bf A}_{n}$ \cite{tdavis1,tdavis2}.

\section{Proposed Algorithm}
In the following, we describe the updating scheme for the inverse of the stiffness matrix
in the case of scalar random fuse model after a fuse has been burnt.
A similar procedure can be applied for central-force and beam models \cite{nukalasiam}.

\par
Let ${\bf A}_{n}$ represent the 
stiffness matrix of the random fuse network system in which $n$ number of 
fuses are either missing (random dilution) or have been burnt during the analysis.
Let us also assume that 
a fuse $ij$ (the $(n+1)^{th}$ fuse)
is burnt when the externally applied voltage is increased gradually.
In the above description, $i$ and $j$ refer to the global degrees of freedom connected 
by the fuse before it is broken. For the scalar random fuse model, the degrees of freedom 
$i$ and $j$ are 
also equivalent to the node $i$ and node $j$ connected by the fuse before it is broken.
The new stiffness matrix ${\bf A}_{n+1}$ of the lattice system after the fuse $ij$ is burnt
is given by
\begin{equation}
{\bf A}_{n+1} = {\bf A}_{n} -k_{ij}~{\bf v} {\bf v}^t \label{A1}
\end{equation}
where
\begin{eqnarray}
{\bf v}^t & = & \left\{ \begin{array}{cccccccccc}
0 & \cdots & 1 & \cdots & -1 & \cdots & 0
\end{array} \right\} \label{Eqv}
\end{eqnarray}
and $k_{ij}$ is the conductance of the fuse $ij$ before it is broken. 
After breaking the 
fuse $ij$, the electrical current in the network is redistributed instantaneously. 
The redistributed current values in the network are calculated by re-solving the 
Kirchhoff equations, i.e., by solving the new set of equations formed by the 
matrix ${\bf A}_{n+1}$. This procedure is very time consuming since a new set of 
equations (inverse of ${\bf A}_{n+1}$ for $n = 0,1,2,\cdots$) 
need to be solved everytime after breaking the $(n+1)^{th}$
fuse. However, significant computational advantages can be gained if the inverse of 
${\bf A}_{n+1}$ is obtained simply by updating the inverse of ${\bf A}_n$. This is achieved 
by using the well known Shermon-Morrison-Woodbury formula for inverting the rank-p update 
of a matrix.
Thus, the inverse ${\bf A}_{n+1}^{-1}$ of Eq. (\ref{A1}) can be expressed as
\begin{eqnarray}
{\bf A}_{n+1}^{-1} & = & \left[{\bf A}_{n}^{-1} + k_{ij}~
\frac{{\bf u} {\bf u}^t}{\left(1 - k_{ij}~{\bf v}^t {\bf u}\right)}\right] \label{A11inv}
\end{eqnarray}
where
\begin{eqnarray}
{\bf u} & = & {\bf A}_{n}^{-1} {\bf v} ~=~ {{\bf A}_{n}^{-1}}_{(i-j)} ~=~
\left[(i^{th} - j^{th})~\mbox{columns of}~{\bf A}_{n}^{-1}\right] 
\label{Equ}
\end{eqnarray}
Hence, the inverse of the stiffness matrix of the lattice system after breaking the
$(n+1)^{th}$
fuse $ij$ is obtained simply by a rank-one update of the inverse of the stiffness 
matrix before the fuse is broken. 
Further, if the inverse of the matrix ${\bf A}_{n}$ is
available explicitly, then the vector ${\bf u}$ can be obtained trivially from 
the $i^{th}$ and $j^{th}$ columns of ${\bf A}_{n}^{-1}$. 
In particular, this implies that if the inverse of the 
matrix ${\bf A}_n$ is available explicitly at 
any stage $n = 0,1,2 \cdots$ of analysis, then the redistributed currents in 
all subsequent stages of analysis 
involving $m = n+1,n+2, \cdots$ burnt fuses can be obtained in a trivial fashion from 
the column vectors of ${\bf A}_n^{-1}$ and the vectors ${\bf u}_p$, where
$p = 1,2,\cdots,(m-n)$. 
However, since the inverse of the 
stiffness matrix ${\bf A}_{n}$ is not usually calculated explicitly, the vector ${\bf u}$ 
is obtained using the already factorized ${\bf A}_n$ matrix through a backsolve 
operation (forward reduction and backward substitution) on the vector ${\bf v}$ 
(Eq. (\ref{Equ})). 

\par
\vskip 0.70em%
\noindent
REMARK 1: Without loss of generality, 
when the fuse that is broken is attached to a {\it constrained/prescribed}
 degree of freedom $j$, the vector ${\bf v}$ is given by 
\begin{eqnarray}
{\bf v}^t & = & \left\{ \begin{array}{cccccccccc}
0 & \cdots & 1 & \cdots & 0
\end{array} \right\} \label{Eqv2}
\end{eqnarray}
and 
\begin{eqnarray}
{\bf u} & = & {{\bf A}_{n}^{-1}}_{(i)} ~=~ 
 \left[i^{th}~\mbox{columns of}~{\bf A}_{n}^{-1}\right] \label{Equ2}
\end{eqnarray}
In the case of periodic boundary conditions, 
consider the case of a broken fuse $jk$ that is attached to a
slave degree of freedom $k$ whose master
degree of freedom is $i$. 
Under these circumstances, the
methodology presented earlier is applicable in a straightforward manner if it is
understood that breaking the fuse $jk$ is equivalent to breaking the fuse $ij$.

\par
\vskip 0.70em%
\noindent
REMARK 2: The load vector ${\bf b}_{n+1}$ will differ from the load vector ${\bf b}_n$ only if the 
$(n+1)^{th}$ broken fuse $ij$ is attached to a prescribed degree of freedom, where a 
constant voltage difference is imposed.
Once again, for presentation purposes, let us assume that $j$ is such a prescribed 
degree of freedom.
Then the load vector ${\bf b}_{n+1}$ is given by
\begin{equation}
{\bf b}_{n+1} = {\bf b}_n + {\bf w} \label{bn1}
\end{equation}
where 
\begin{eqnarray}
{\bf w}^t = k_{ij}~\left\{\begin{array}{cccccccccc}
0 & 0 & \cdots & -1 & \cdots & 0 & 0 
\end{array} \right\}
\end{eqnarray}
If neither $i$ nor $j$ is a prescribed degree of freedom, then ${\bf w} = {\bf 0}$.

\par
Before breaking the $(n+1)^{th}$ fuse, the solution vector ${\bf x}_n$ 
is obtained by solving the Kirchhoff equations 
\begin{equation}
{\bf A}_n {\bf x}_n = {\bf b}_n \label{Kchn}
\end{equation}
After breaking the $(n+1)^{th}$ fuse that connects the $i^{th}$ and $j^{th}$ degrees of freedom, 
the updated solution vector ${\bf x}_{n+1}$ is obtained by 
solving the new set of Kirchhoff equations 
\begin{equation}
{\bf A}_{n+1} {\bf x}_{n+1} = {\bf b}_{n+1} \label{Kchnp1}
\end{equation}
Substituting Eqs. (\ref{A11inv},\ref{bn1}) and (\ref{Kchn}) into the solution of 
Eq. (\ref{Kchnp1}) and simplifying the result, we have
\begin{eqnarray}
{\bf x}_{n+1} & = & {\bf A}_{n+1}^{-1} {\bf b}_{n+1} \nonumber \\
& = & \left[{\bf A}_{n}^{-1} + {k_{ij}}~
\frac{{\bf u} {\bf u}^t}{\left(1 - {k_{ij}}~{\bf v}^t {\bf u}\right)}\right]~\left({\bf b}_n + {\bf w}\right) \nonumber \\
& = & {\bf x}_n ~+~ \beta~{\bf u} \label{xn1}
\end{eqnarray}
where
\begin{eqnarray}
\beta & = & \alpha~\left({\bf u}^t {\bf b}_{n+1}\right) - {k_{ij}} ~~~~\mbox{if 
$i$ or $j$ is prescribed} \nonumber \\
& = & \alpha~\left({\bf u}^t {\bf b}_{n+1}\right) ~~~~\mbox{otherwise}
\end{eqnarray}
and 
\begin{eqnarray}
\alpha & = & \frac{k_{ij}}{\left(1 - {k_{ij}}~{\bf v}^t {\bf u}\right)} 
\end{eqnarray}

\par
The only unknown in Eq. (\ref{xn1}) is the column vector ${\bf u}$,
which can be obtained through a backsolve operation using either Eq. (\ref{Equ}) or 
Eq. (\ref{Equ2}). Furthermore, it is not necessary to explicitly 
assemble the matrix ${\bf A}_n$ and perform factorization to do the 
backsolve operation. Instead, we can use the already
factorized matrix ${\bf A}_{m}$ to obtain the vector ${\bf u}$.
In the above description, $m < n$ and denotes the latest broken bond at 
which the factorization ${\bf L}_{m}$ of ${\bf A}_{m}$ is available.
To see this clearly, let us first decompose
the matrix ${\bf A}_{n}^{-1}$ into ${\bf A}_{m}^{-1}$ and a matrix ${\bf C}$ such that
\begin{equation}
{\bf A}_{n}^{-1} = {\bf A}_{m}^{-1} + {\bf C} \label{dcn}
\end{equation}
where
\begin{eqnarray}
{\bf C} & = & \sum_{l=1}^{p = (n-m)} {k_{l}}~
\frac{{\bf u}_l {\bf u}_l^t}{\left(1 - {k_{l}}~{\bf v}_l^t {\bf u}_l\right)} \label{Cn}
\end{eqnarray}
Due to the amount of the storage requirement $(\sim O(n_{dof}^{2}))$, and the computational cost 
associated in evaluating the Eq. (\ref{Cn}) $(\sim O(n_{dof}^{2}))$, the matrix 
${\bf C}$ is never explicitly calculated or stored. Instead, the vectors ${\bf u}_l$ 
for $l = 1,2,\cdots,(n-m)$ are stored, and the $(j^{th} - i^{th})$ column of ${\bf C}$ 
is evaluated as
\begin{eqnarray}
{{\bf C}}_{(j-i)} & = & \sum_{l=1}^{p = (n-m)} {k_{l}}~ 
\frac{({{\bf u}_l}_j - {{\bf u}_l}_i)}{\left(1 - {k_{l}}~{\bf v}_l^t {\bf u}_l\right)}~{\bf u}_l \label{Cnv}
\end{eqnarray}
where ${{\bf u}_l}_i$ and ${{\bf u}_l}_j$ refer to the $i^{th}$ and $j^{th}$ components
of the vector ${\bf u}_l$.
Equation (\ref{Cnv}) 
reduces the storage and computational cost to $(\sim O(p ~n_{dof}))$ and 
$(\sim O(p ~n_{dof}))$ operations, respectively. 
Even with this modification, 
the storage and computational requirements can become prohibitively expensive
as the number of updates, $p$, increases, 
and hence it is necessary to limit the maximum number of vector updates
between two successive factorizations 
to a certain $maxupd$. That is, it is necessary to perform or update the 
factorization of the stiffness matrix ${\bf A}$ at regular intervals. 

\par
Instead of re-factorizing the stiffness matrix ${\bf A}$ after 
every $maxupd$ steps, it is more effective to update the factorization ${\bf L}_m$ using the 
multiple-rank sparse Cholesky factorization update algorithm \cite{tdavis1,tdavis2}.
This multiple-rank update of ${\bf L}_m$ to
obtain the new factorization ${\bf L}_{n+1}$, after breaking the $(n+1)^{th}$ fuse, is
computationally cheaper compared with the direct factorization ${\bf L}_{n+1}$
of the new stiffness matrix ${\bf A}_{n+1}$ \cite{tdavis1,tdavis2}. 
We use the multiple-rank downdate algorithm presented in \cite{tdavis1,tdavis2} 
to obtain the new Cholesky factorization ${\bf L}_{n+1}$ from the existing 
Cholesky factor ${\bf L}_m$.
The multiple-rank downdate algorithm \cite{tdavis1,tdavis2} is based on the analysis and 
manipulation of the underlying graph structure of the stiffness matrix ${\bf A}$
and on the methodology presented in Gill et al. \cite{gill74,gill75} for modifying a 
dense Cholesky factorization. This algorithm incorporates the change in the 
sparsity pattern of ${\bf L}$ and is optimal in the sense that 
the computational time required is proportional to the number of changing non-zero 
entries in ${\bf L}$.
In particular, since the breaking of fuses is equivalent to removing the 
edges in the underlying graph structure of stiffness matrix ${\bf A}$, the 
new sparsity pattern of the modified ${\bf L}$ must be a subset of the sparsity pattern 
of the original ${\bf L}$. Denoting the sparsity pattern of ${\bf L}$ by ${\mathcal L}$, we have
\begin{eqnarray}
{\mathcal L}_{m} \supseteq {\mathcal L}_{n} ~~~\forall ~m < n
\end{eqnarray}
Therefore, we can even use the modified dense Cholesky factorization update (algorithm 5 in the 
reference Davis et al. \cite{tdavis1}) and work only on the non-zero entries in ${\bf L}$. 
Furthermore, since the changing non-zero entries in ${\bf L}$ depend on the $i^{th}$ and 
$j^{th}$ degrees of freedom of the fuse $ij$ that is broken, it is only necessary to 
modify the non-zero elements of a submatrix of ${\bf L}$. 

\par
The multiple-rank update of the 
sparse Cholesky factorization is computationally superior to an equivalent series of 
rank-one updates since the multiple-rank update makes one pass through ${\bf L}$ 
in computing the new entries, while a series of rank-one updates require multiple passes
through ${\bf L}$ \cite{tdavis2}. 
The multiple-rank update algorithm updates the 
Cholesky factorization ${\bf L}_m$ of the matrix ${\bf A}_m$ to ${\bf L}_{n+1}$ of the 
new matrix ${\bf A}_{n+1}$, where ${\bf A}_{n+1} = {\bf A}_m + \sigma {\bf Y}{\bf Y}^t$, 
and ${\bf Y}$ represents a $n_{dof} \times p$ rank-p matrix. 
The computational cost involved in breaking the $(n+1)^{th}$ fuse $ij$ 
is simply a backsolve operation $(O(nnz({\bf L}_m)))$
on a load vector given by Eq. (\ref{Eqv}) using the already
factored matrix ${\bf A}_m$, $(n+2-m)$ vector updates, and one vector inner product.

\par
The optimum number of steps between successive factorizations of the matrix ${\bf A}$ 
is determined by minimizing the computatioal cpu time required for the 
entire analysis. Let $t_{fac}$ and $t_{upd}$ denote the average cpu time required 
for performing/updating the 
factorization ${\bf A}_m$ and the average cpu time required for a single 
rank-1 update of the solution ${\bf \tilde{u}}_{(n+1-m)}$, respectively. Note that the evaluation of 
${\bf \tilde{u}}_{(n+1-m)}$ requires $(n-m)$ vector updates. Let the 
estimated number of steps for the lattice system failure be $n_{steps}$. 
Then, the total cpu time required for solving the linear system of equations until the  
lattice system failure is given by
\begin{eqnarray}
\Psi & = & n_{fac} t_{fac} + \sum \frac{\left(n_{steps} - n_{fac}\right)}{n_{fac}} ~t_{upd} 
\nonumber \\
& = & n_{fac} t_{fac} + \frac{1}{2}~\frac{\left(n_{steps} - n_{fac}\right)}{n_{fac}} ~\frac{n_{steps}}{n_{fac}} ~t_{upd}
\end{eqnarray}
where $n_{fac}$ denotes the number of factorization until lattice system failure. The 
optimum number of factorizations, $n_{opt-fac}$, for the entire analysis is obtained by 
minimizing the function $\Psi$. The maximum number of vector updates, $maxupd$, between 
successive factorizations is estimated as
\begin{equation}
maxupd = \frac{\left(n_{steps} - n_{opt-fac}\right)}{n_{opt-fac}}
\end{equation}

\section{Numerical Simulation Results}
In the following, we consider two alternate forms of the algorithm presented in this 
paper. These two solver types are
\begin{itemize}
\item {\it Solver Type A}: Given the factorization ${\bf L}_m$ of 
${\bf A}_m$, we use rank-1 sparse Cholesky update/downdate \cite{tdavis1} 
to update the factorization ${\bf L}_{n+1}$ ($O(nnz({\bf L}_n)$) for all subsequent values of $n = m,m+1, \cdots$.
Once the factorization ${\bf L}_{n+1}$ of ${\bf A}_{n+1}$ is obtained, the solution vector 
${\bf x}_{n+1}$ is obtained by a backsolve operation ($O(nnz({\bf L}_{n+1})$).

\item {\it Solver Type B}: Given the factorization ${\bf L}_m$ of 
${\bf A}_m$, the algorithm evaluates the new solution vector 
${\bf x}_{n+1}$, after the $(n+1)^{th}$ fuse is burnt, using Eq. (\ref{xn1}) 
($O(nnz({\bf L}_m)$ + $(n+2-m)$ vector updates). 
Instead of refactorizing the matrix after $maxupd$ steps, we use 
rank-p sparse Cholesky update/downdate \cite{tdavis2} to obtain the factorization 
${\bf L}_{m+maxupd}$ of the matrix ${\bf A}_{m+maxupd}$ ($O(nnz({\bf L}_m)$).
\end{itemize}
The above two algorithms are benchmarked against the 
PCG iterative solvers, in which {\it optimal} \cite{tchan88,chan89,chan921,chan96} 
circulant matrices are used as preconditioners to the Laplacian operator (Kirchhoff equations). 
The Fourier accelerated PCG presented in \cite{batrouni86,batrouni88,batrouni98}
is not optimal in the sense described in \cite{tchan88,chan89,chan921,chan96}, and 
hence it is expected to take more number of CG iterations compared with the {\it optimal} 
circulant preconditioners. 

\par
In the numerical simulations using solver types A and B,
the supernodal Cholesky factorization is performed using the TAUCS solver library
({\it http://www.tau.ac.il/~stoledo/taucs}).
In these simulations, 
the maximum number of vector updates, $maxupd$, is chosen to be a constant 
for a given lattice size. We choose $maxupd = 25$ for $L = \{4, 8, 16, 24, 32\}$, 
$maxupd = 50$ for $L = 64$, and $maxupd = 100$ for $L = \{128, 256, 512\}$. For 
$L = 512$, $maxupd$ is limited to 100 due to memory constraints. 
By keeping the $maxupd$ value constant, it is possible to realistically 
compare the computational cost associated with different solver types. Moreover, the 
relative cpu times taken by these algorithms remains the same even when the simulations 
are performed on different platforms. 

\par
Tables 1 and 2 present the cpu and wall-clock times taken for one configuration (simulation) 
using the solver types A and B, respectively. These tables also indicate the 
number of configurations, $N_{config}$, over which ensemble averaging of the 
numerical results is performed. The cpu and wall-clock times taken by the 
{\it optimal} circulant matrix preconditioned iterative solver is presented in Tables 3. 
For iterative solvers, the number of iterations presented in Tables 3 denote 
the average number of total iterations taken to break one intact lattice configuration 
until it falls apart.

\par
Based on the results presented in Tables 1-3, it is clear that for modeling the 
breakdown of disordered media as in starting with an intact lattice and successive 
breaking of bonds until the lattice system falls apart, the solver types A and B 
based on direct solvers are superior to the Fourier accelerated iterative PCG solver techniques. 
It should be noted that for larger lattice 
systems, limitations on the available memory of the processor may 
decrease the allowable $maxupd$ value, 
as in the case of $L = 512$ using solver type B. However, this is not a concern 
for simulations performed using solver type A. 

\par
\begin{figure}[hbtp]
\centerline{\includegraphics[width=12cm]{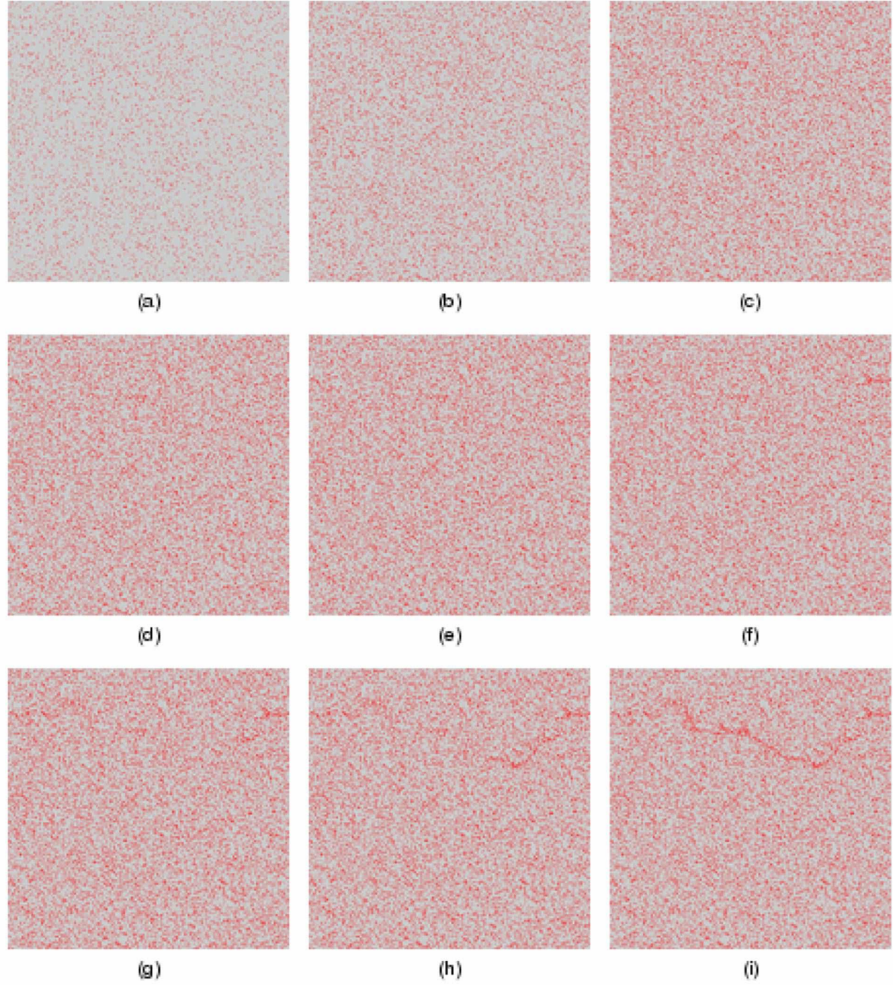}}
\caption{Snapshots of damage in a typical triangular lattice system of size $L = 512$. Number of broken bonds at the peak load and at failure are 83995 and 89100, respectively. (a)-(i) represent the snapshots of damage after breaking $n_b$ number of bonds. (a) $n_b = 25000$ (b) $n_b = 50000$ (c) $n_b = 75000$ (d) $n_b = 80000$ (e) $n_b = 83995$ (peak load) (f) $n_b = 86000$ (g) $n_b = 87000$ (h) $n_b = 88000$ (i) $n_b = 89100$ (failure)}
\label{fig1a}
\end{figure}

Using the solver type A, we have performed numerical simulations on two-dimensional 
triangular and diamond (square lattice inclined at
45 degrees between the bus bars) lattice networks. 
Table 4 presents the number of broken bonds at peak load, $n_p$, and
at fracture, $n_f$, for each of the lattice sizes considered. In addition, 
Table 4 also presents the number of configurations, $N_{config}$,
over which statistical averaging is
performed for different lattice sizes. The numerical results presented in 
Tables 1-3 are performed on a single processor of 
{\it Cheetah} (27 Regatta nodes with thirty two 1.3 GHz Power4 processors each), 
the eighth fastest supercomputer in the world
({\it http://www.ccs.ornl.gov}). However, the numerical simulation results presented in 
Table 4 are performed on {\it Eagle} (184 nodes with four 375 MHz Power3-II processors) 
supercomputer at the Oak Ridge National Laboratory to run simulations simultaneously 
on more number of processors.
Figure \ref{fig1a} presents the snapshots of progressive damage
evolution for the case of a broadly distributed random thresholds model problem
in a triangular lattice system of size $L = 512$.

\section{Conclusions}
The paper presents an algorithm based on 
rank-one update of the inverse of the stiffness matrix and the multiple-rank downdating of the 
sparse Cholesky factorization for simulating fracture and damage 
evolution in disordered quasi-brittle materials using discrete lattice networks. 
Using the proposed algorithm, the average computational cost associated with 
breaking a bond reduces to the same order as that of a simple backsolve (forward elimination and 
backward substitution) operation using the already LU factored matrix. 
This algorithm based on direct solver techniques eliminates 
critical slowing down observed in fracture simulations 
using the conventional iterative schemes. 
Numerical simulations on random resistor networks demonstrate that 
the present algorithm is computationally superior to the commonly used 
Fourier accelerated preconditioned conjugate gradient iterative solver.

\par
For analysis of fracture
simulations using discrete lattice networks, ensemble averaging of
numerical results is necessary to obtain a realistic representation of
the lattice system response. In this regard, 
for very large lattice systems with large number of system of equations, this 
methodology is especially advantageous as the LU factorization of the system of 
equations can be performed using a parallel implementation on multiple processors.
Subsequently, this factored LU decomposition can then be distributed to each of the 
processors to continue with independent fracture simulations that only 
require less intensive backsolve operations.

\par
\vskip 1.00em%
\noindent
{\bf Acknowledgment} \\
This research is sponsored by the Mathematical, Information and Computational Sciences
Division, Office of Advanced Scientific Computing Research, U.S. Department of Energy under
contract number DE-AC05-00OR22725 with UT-Battelle, LLC. The first author wishes to thank
Ed F. D'Azevedo for many helpful discussions and excellent suggestions.

\newpage

\bibliography{bfgs_direct}
\bibliographystyle{unsrt}

\newpage

\begin{table}[hbtp]
  \leavevmode
  \begin{center}
  \caption{Computational cost associated with solver type A}
  \vspace*{1ex}
  \begin{tabular}{|c|c|c|c|}\hline
  Size  & CPU(sec) & Wall(sec) & Simulations \\
  \hline
 32 & 0.592 & 0.687 & 20000 \\
 64 & 10.72 & 11.26 & 4000 \\
128 & 212.2 & 214.9 & 800 \\
256 & 5647 & 5662 & 96 \\
512 & 93779 & 96515 & 16 \\
  \hline
  \end{tabular}
  \label{table1}
  \end{center}
\end{table}

\begin{table}[hbtp]
  \leavevmode
  \begin{center}
  \caption{Computational cost associated with solver type B}
  \vspace*{1ex}
  \begin{tabular}{|c|c|c|c|}\hline
  Size  & CPU(sec) & Wall(sec) & Simulations \\
  \hline
 32 & 0.543 & 0.633 & 20000 \\
 64 & 11.15 & 12.01 & 4000 \\
128 & 211.5 & 214.1 & 800 \\
256 & 6413 & 6701 & 96 \\
  \hline
  \end{tabular}
  \label{table2}
  \end{center}
\end{table}

\begin{table}[hbtp]
  \leavevmode
  \begin{center}
  \caption{Computational cost associated with {\it optimal} circulant PCG}
  \vspace*{1ex}
  \begin{tabular}{|c|c|c|c|c|}\hline
  Size  & CPU(sec) & Wall(sec) & Iterations & Simulations \\
  \hline
 32 & 11.66 & 12.26 & 25469 & 20000 \\
 64 & 173.6 & 178.8 & 120570 & 1600 \\
128 & 7473 & 7725 & 622140 & 128 \\
  \hline
  \end{tabular}
  \label{table3}
  \end{center}
\end{table}

\newpage

\begin{table}[hbtp]
  \leavevmode
  \begin{center}
  \caption{Number of broken bonds at peak and at failure}
  \vspace*{1ex}
  \begin{tabular}{|c|c|c|c|c|c|c|}\hline
  L  & $N_{config}$ & time & \multicolumn{2}{c|}{Triangular} & \multicolumn{2}{c|}{Diamond}\\\cline{4-7}
     & & (seconds) & $n_p$ & $n_f$ & $n_p$ & $n_f$\\
  \hline
  4 & 50000 & 0.002 & 13 & 19 & 9 & 14 \\
  8 & 50000 & 0.006 & 41 & 54 & 26 & 37 \\
 16 & 50000 & 0.042 & 134 & 168 & 80 & 107 \\
 24 & 50000 & 0.186 & 276 & 335 & 161 & 208 \\
 32 & 50000 & 0.592 & 465 & 554 & 268 & 337 \\
 64 & 50000 & 10.72 & 1662 & 1911 & 942 & 1126 \\
128 & 12000 & 212.2 & 6068 & 6766 & 3406 & 3901 \\
256 & 1200 & 5647 & 22572 & 24474 & 12571 & 13846 \\
512 & 200 & 93779 & 84487 & 89595 &  &  \\
  \hline
  \end{tabular}
  \label{table4}
  \end{center}
\end{table}

\newpage

\end{document}